\documentclass[cits]{PoS}
\usepackage{graphicx}
\usepackage{amstext}

\def\n{\noindent}
\def\ie{\emph{$i.e.$ }}
\def\eg{\emph{$e.g.$ }}

\def\suN{SU$(N)$ }

\PoS{PoS(LAT2005)282}

\title{Phase transitions and non-analyticities in large N$_\text{c}$ gauge
theories}

\ShortTitle{Phase transitions and non-analyticities in large N$_\text{c}$
gauge theories}

\author{Francis Bursa\\
        University of Oxford\\
        E-mail: \email{bursa@thphys.ox.ac.uk}}

\author{Michael Teper\\
        University of Oxford\\
        E-mail: \email{teper@thphys.ox.ac.uk}}

\author{\speaker{Helvio Vairinhos}\\
        University of Oxford\\
        E-mail: \email{helvio@thphys.ox.ac.uk}}

\abstract{We investigate numerically various phase transitions and
non-analyticities at large $N$ using both twisted Eguchi-Kawai
space-time reduction and the standard Wilson theory.}

\FullConference{XXIIIrd International Symposium on Lattice Field
Theory\\25-30 July 2005\\Trinity College, Dublin, Ireland}

\begin{document}

\section{Introduction}\label{sec:Intro}

Despite the considerable simplifications that occur when one takes
the 't Hooft limit of pure \suN gauge theories, the resulting
theory is complex enough to allow a very rich and interesting
physics. The phase structure of large $N$ lattice gauge theory on
the 4-torus is a good example of this richness. Several
transitions were identified and studied in the last few years, \eg
the `bulk' transition \cite{Campostrini98}, the finite temperature
deconfining transition \cite{LTW02} and the recently discovered
transitions between phases with a different number of broken
directions \cite{KNN03}; a recently conjectured non-analyticity
associated with the physical size of Wilson loops might also join
this list \cite{KNN03}.

The purpose of this paper is to present some preliminary results 
of a numerical study on the phase structure of 4D Euclidian pure 
SU$(\infty)$ gauge theory.

\section{TEK model}\label{sec:TEK}

A very useful alternative to the usual approach of large $N$
extrapolations in lattice gauge theory \cite{Teper.LAT05} is the
idea of space-time reduction \cite{nEK}. Due to factorization, the
Schwinger-Dyson (loop) equations for infinite- and finite-volume
lattice gauge theories coincide in the large $N$ limit. This means
that space-time degrees of freedom become spurious at large $N$,
so that the properties of the SU$(\infty)$ gauge theory can be
analyzed with very small lattices (\emph{in extremis} with an
one-point lattice), as long as the gauge group is large enough.
This trade between space-time and color degrees of freedom allows
us to analyze very large gauge groups without great computational
effort.

The \emph{twisted Eguchi-Kawai} (TEK) model is a particular case
of a reduced model, consisting of a \suN gauge theory defined on a
twisted $1^4$ lattice. It should reproduce a \suN gauge theory
defined on a periodic $L^4$ lattice with the standard Wilson
action up to $O(\frac{1}{N^2})$ corrections. The TEK action is given by
    \begin{equation}\label{eq:action}
    S_\mathrm{TEK}(U)=bN\sum^4_{\mu>\nu}\mathrm{Tr}\,(Z_{\mu\nu}
    U_\mu U_\nu U_\mu^\dag U_\nu^\dag+\mathrm{h.c.}),
    \end{equation}
\n where $b=\frac{1}{g^2 N}$ is the inverse 't Hooft coupling,
$Z_{\mu\nu}=\mathrm{exp}\left(\frac{2\pi i}{N}n_{\mu\nu}\right)$
is the twist, $n_{\mu\nu}$ is an antisymmetric tensor chosen to be
$n_{\mu\nu}=L$ for all $\mu>\nu$, and $L^4=N^2$ \cite{AO82}. In
addition to the usual gauge symmetry, $U_\mu\mapsto\Omega
U_\mu\Omega^\dag$ ($\Omega\in\mathrm{SU}(N)$), the TEK action
(\ref{eq:action}) also has a global $Z^4_N$ symmetry,
$U_\mu\mapsto z_\mu U_\mu$ ($z_\mu\in Z_N$), which is unbroken for
all values of the coupling $b$. The observables of the TEK model
are obtained from the ones in the standard Wilson theory by making
the substitution $U_\mu(x)\mapsto D(x)U_\mu D(x)^\dag$, where
$D(x)=\prod^4_{\mu=1}\Gamma^{x_\mu}_\mu$ and $\Gamma_\mu$ are the
vacuum matrices that extremize eqn(\ref{eq:action}). In
particular, the reduced Wilson and Polyakov loops are of the form,
respectively,
    \begin{equation}\label{eq:Wilson}
    W(I,J)=\frac{1}{N}\mathrm{Tr}\,Z^{IJ}_{\mu\nu}U^I_\mu U^J_\nu
    {U_\mu^\dag}^I {U_\nu^\dag}^J,
    \end{equation}
    \begin{equation}\label{eq:Polyakov}
    P_\mu=\frac{1}{N}\mathrm{Tr}\,U_{\mu}^L\;.
    \end{equation}
The equivalence between the standard Wilson theory and the TEK
model states that the expectation values of gauge-invariant
observables in these models should coincide up to
$O(\frac{1}{N^2})$ corrections:
    \begin{equation}\label{eq:equivalence}
    \langle\mathcal{O}[U]\rangle_{\mathrm{W}}=\langle\mathcal{O}
    [DUD^\dag]\rangle_{\mathrm{TEK}}+O\left(\frac{1}{N^2} \right).
    \end{equation}

\section{Wilson loop non-analyticity}\label{sec:Wilson}

\begin{figure}[t]
    \begin{center}
    \leavevmode\scalebox{0.85}{\input{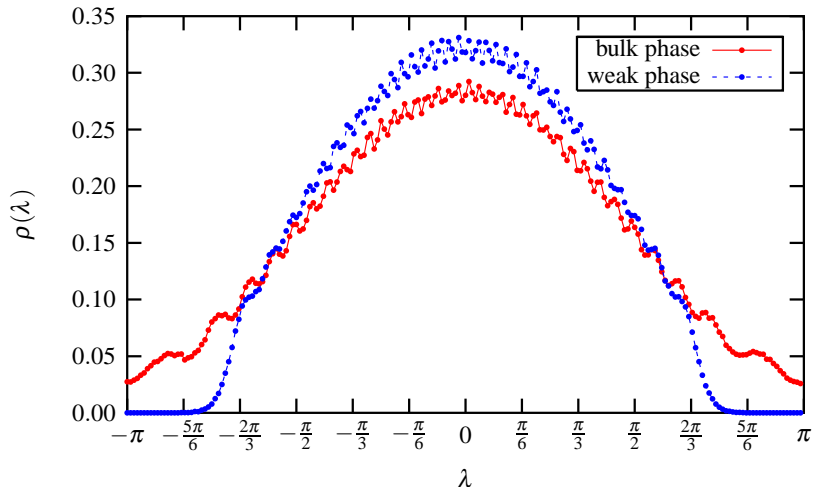}}
    \end{center}
    \vspace{-7mm}\caption{Effect of the `bulk'
    transition on the eigenvalue spectrum of the plaquette, 
    simulated in the SU(36) TEK model at fixed coupling $b=0.3540$: 
    a spectral gap forms in the `weak-coupling' phase.}
    \label{fig:eigspec}
    \end{figure}

The `bulk' transition is a lattice artifact that affects 4D
lattice gauge theories with gauge group SU$(N\geq 5)$, occurring
at a critical coupling $b_c\approx 0.36$ \cite{Campostrini98}. It
is strongly first order and manifests itself by a non-analytic
change in the eigenvalue spectrum of the plaquette: in the `bulk'
phase the eigenvalue spectrum is spread over the whole unit 
circle\footnote{The eigenvalues of \suN matrices are just phases 
of the form $e^{i\lambda}$}, while in the `weak-coupling' phase a 
spectral gap forms around $\lambda=-\pi$ (Fig.\ref{fig:eigspec}).
Recently \cite{KNN03}, it was conjectured that a similar
transition might occur for larger Wilson loops that would affect
their eigenvalue spectrum in a similar way: at $N=\infty$ a
spectral gap would form (non-analytically) around $\lambda=-\pi$,
for a specific critical value of the coupling that would scale
with the \emph{physical size} of the Wilson loop. We looked for
this transition in numerical simulations performed in both the
standard Wilson theory and the TEK model.

\subsection{Method}

The method used to detect the formation of a gap in the eigenvalue 
spectrum of Wilson loops consisted in measuring the changes in the 
fluctuations of their individual eigenvalues, $\lambda_i$. For that 
purpose we considered the following ratio of correlation functions:
    \begin{equation}\label{eq:correlation}
    R=\frac{\langle\lambda^2_1\rangle-\langle
    \lambda_1\rangle^2}{\langle
    \lambda^2_\frac{N}{2}\rangle-\langle
    \lambda_\frac{N}{2}\rangle^2}
    \end{equation}
This quantity is gauge-invariant, because it only depends on the
eigenvalues of the Wilson loop. It corresponds to the fluctuations 
of the eigenvalue closest to $\lambda=-\pi$, $\lambda_1$, 
normalized by the fluctuations of the one closest to $\lambda=0$, 
$\lambda_{\frac{N}{2}}$. For small $b$ the eigenvalue spectrum is 
spread over the whole interval $[-\pi,\pi]$, without gaps, which 
leaves enough room for the individual eigenvalues to fluctuate; for 
large $b$ there is a gap, so the eigenvalues are squeezed and have 
less room to fluctuate (Fig.\ref{fig:eigspec}). The fluctuations 
of the central eigenvalues like $\lambda_{\frac{N}{2}}$ are in 
general more affected by this squeezing (\ie decrease faster) then 
the fluctuations of the outer ones, which results in an increasing 
of the ratio $R$ across the transition from a gapless to a gapped 
phase. Therefore, if the formation of a gap in the eigenvalue 
spectrum of Wilson loops is to be a non-analytical process, we also 
expect the ratio $R$ to show a sudden jump at the same critical 
value of the coupling $b$.

\subsection{Results}

\begin{figure}[t]
    \begin{center}
    \leavevmode\scalebox{0.85}{\input{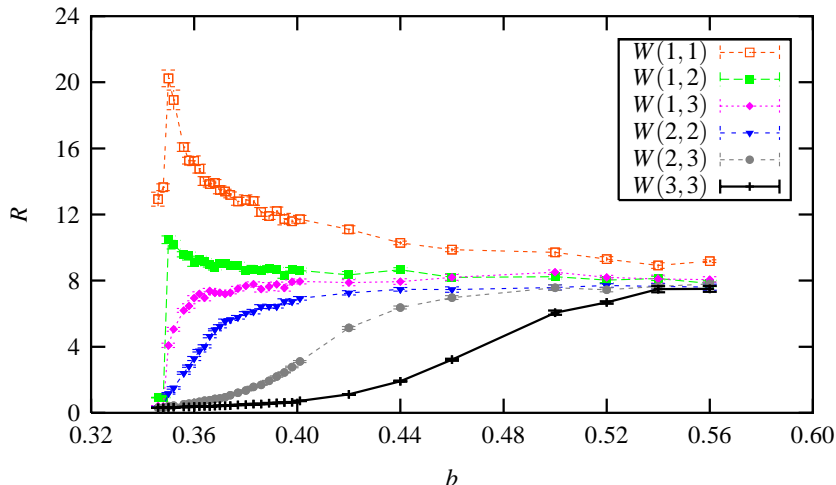}}
    \end{center}
    \vspace{-7mm}\caption{Ratio of correlations $R$ vs coupling $b$, 
    in the SU(81) TEK model for Wilson loops of several sizes. There
    is a clear first order `bulk' transition for the plaquette,
    $W(1,1)$, as expected. Transitions for larger loops also exist, 
    but they are very smooth.}
    \label{fig:correlations:fixed-N}
    \end{figure}

We performed Monte Carlo simulations to calculate the ratio $R$ of
eqn(\ref{eq:correlation}). We simulated the TEK model with gauge
groups $N=25,36,49,64,81$ (which correspond to effective lattices
of size $L=5,6,7,8,9$, respectively); we also simulated the
standard Wilson theory with gauge groups $N=6,12$ on a $6^4$
lattice. The graph in Fig.\ref{fig:correlations:fixed-N} shows the
change of $R$ with the coupling $b$ for Wilson loops of several
sizes and in the SU(81) TEK model.

\begin{figure}[t]
    \begin{center}
    \leavevmode\scalebox{0.85}{\input{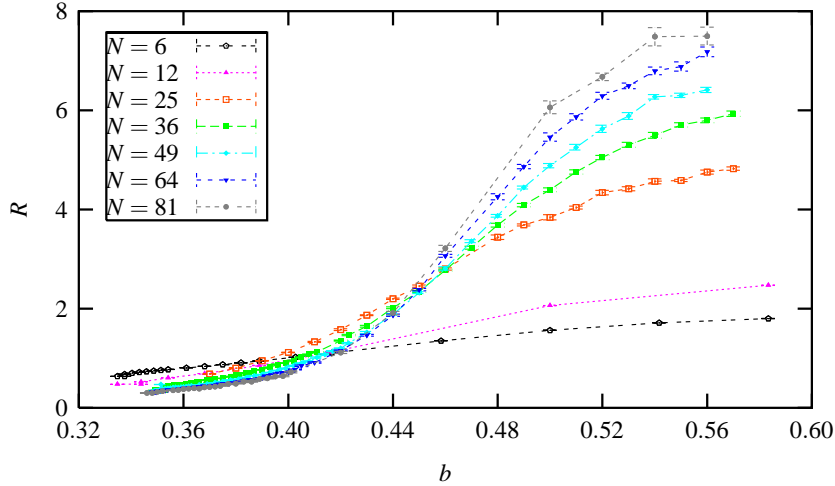}}
    \end{center}
    \vspace{-7mm}\caption{Ratio of correlations $R$ vs coupling $b$,
    for Wilson loops of fixed lattice size, $W(3,3)$, and for
    several gauge groups. The transition shows a slow evolution
    towards a sharper transition for increasing $N$, but the data
    doesn't allow us to reach a definitive conclusion.}
    \label{fig:correlations:fixed-size}
    \end{figure}

The `bulk' transition, at $b\approx 0.35$, has a very clear
non-analytic effect on $R$ for all Wilson loops (especially for
the plaquette). For the transitions at larger values of $b$
associated with the formation of a gap in the eigenvalue spectra
of large Wilson loops, however, the situation is different: the
transitions can be easily seen to exist
(Fig.\ref{fig:correlations:fixed-N}), but they are very smooth.
Justifications for this smooth behavior could be 1) $R$ is not a
good candidate for an order parameter of the transition, 2) the
transition exists at $N=\infty$, but has large $\frac{1}{N}$ 
corrections, or 3) there isn't a non-analyticity at all, only a 
smooth crossover (or it might not even scale correctly with the 
physical size of the Wilson loops). The answer to these questions 
is still unknown to us and are being checked. All we can say from 
these results is that there are in fact transitions, but also that 
it is very hard to tell how they evolve with $N$. If one compares how 
$R$ changes with $b$ for several gauge groups and for a Wilson loop of 
fixed lattice size, $W(3,3)$ for example, we notice a very slow 
evolution with increasing $N$ towards a sharper transition 
(Fig.\ref{fig:correlations:fixed-size}). This tendency could result in 
a non-analyticity at large $N$, but the data doesn't allow us to reach 
a definitive conclusion.

\section{Polyakov loop transitions}\label{sec:Polyakov}

The Polyakov loop is a gauge-invariant observable that winds
around a non-contractible loop of the 4-torus. In the standard
Wilson picture, its expectation value $\langle
P_\mu\rangle_\mathrm{W}$ monitors the deconfining transition in
pure \suN gauge theories, serving as its order parameter. This
transition is associated with the spontaneous breaking of a global
$Z_N$ symmetry.

The symmetry group, $Z_N$, acts on Polyakov loops in the following
way:
    \begin{equation}\label{eq:Z(N)symmetry}
    P_\mu\mapsto z\,P_\mu\;\;\;\;(z\in Z_N).
    \end{equation}
This action consists in multiplying all links in a given layer of
the lattice, $\{U_\mu(x)|\mu$=constant, $x^\mu$=constant$\}$,
with an element $z\in Z_N$; consequently, all Polyakov loops in
the $\mu$-direction come multiplied by $z$. In the confined phase
the $Z_N$ symmetry is realized and $\langle
P_\mu\rangle_\mathrm{W}$ is forced to be zero, while in the deconfined
phase that symmetry is spontaneously broken and there are $N$
possible deconfined phases, $\langle
P_\mu\rangle_\mathrm{W}\propto \mathrm{exp}(\frac{2\pi i k}{N})\in
Z_N$, $k=0,\ldots,N-1$.

In the TEK model, however, the gauge fields live on a $1^4$
lattice (there is only one layer of links in each direction) and
the reduced Polyakov loops are defined as the $L$th power of the
same link variable, eqn(\ref{eq:Polyakov}). Therefore, the
symmetry group that acts on the reduced Polyakov loops is $Z_L$
and not $Z_N$, due to the $Z^4_N$ symmetry of the TEK action:
    \begin{equation}\label{eq:Z(L)symmetry}
    U_\mu\mapsto z_\mu\,U_\mu\;\;\Rightarrow\;\;P_\mu\mapsto
    z^L_\mu\,P_\mu\;\;\;\;(z_\mu\in Z_N\;\;\Rightarrow\;\;z^L_\mu
    \in Z_L).
    \end{equation}
This $Z_L$ symmetry of the reduced Polyakov loops breaks at a
given critical coupling, which in all the numerical simulations
performed coincided with the `bulk' transition. The corresponding
`confined' phase is characterized by $\langle
P_\mu\rangle_{\mathrm{TEK}}=0$, just like in the standard Wilson
theory, while in the `deconfined' phase (where the $Z_L$ symmetry
is broken) there are \emph{only} $L=\sqrt N$ possible deconfined
phases, $\langle P_\mu\rangle_\mathrm{TEK}\propto
\mathrm{exp}(\frac{2\pi i k}{L})\in Z_L$, $k=0,\ldots,L-1$. This
property reveals an explicit difference between the TEK reduced
models and their equivalent standard Wilson theory: they haven't
the same number of Polyakov loop phases, contrary to what one might
na\"ively expect from eqn(\ref{eq:equivalence}). This observation 
might lead to the question of whether the usual argument of 
equivalence between TEK models and standard Wilson theory using 
lattice Schwinger-Dyson equations at large $N$ also applies to
non-contractible loops, and consequently if a study of the
deconfining transition using reduced Polyakov loops is valid.

\section{Conclusions}\label{sec:Conclusion}

We presented some preliminary  results of a numerical study on the 
phase structure of large $N$ gauge theories on the 4-torus. We 
showed evidence of the existence of transitions associated with a 
gap formation in the eigenvalue spectrum of large Wilson loops. 
These transitions, however, are very smooth and their evolution 
with increasing $N$ does not allow conclusions about their nature 
at $N=\infty$. To conclusively infer from numerical calculations 
the existence (or not) of the Wilson loop non-analyticity discussed 
in section \ref{sec:Wilson} we might need to simulate larger gauge 
groups ($N>81$), for which the TEK model is the only relevant and 
practicable method.

We also checked the behavior of reduced Polyakov loops in the TEK
model. It can be easily seen from symmetry arguments that reduced
Polyakov loops behave differently from their standard Wilson
theory equivalents, contrary to what one might na\"ively expect from
eqn(\ref{eq:equivalence}). Therefore, the transitions of reduced
Polyakov loops might not be a good tool for studying the large $N$
deconfining transition.

\vspace{5mm}
\section*{Acknowledgements}
Our lattice calculations were carried out on PPARC and EPSRC
funded computers in Oxford Theoretical Physics. FB acknowledges
the support of a PPARC graduate studentship. HV is supported by
FCT (Portugal) under the grant SFRH/BD/12923/2003.


\begin{thebibliography}{99}

\bibitem{Campostrini98}
M. Campostrini,
\emph{Nucl. Phys. Proc. Suppl.}  {\bf 73} (1999) 724.

\bibitem{LTW02}
B. Lucini, M. Teper and U. Wenger, \emph{Phys. Lett.} {\bf B545}
(2002) 197 [{\tt hep-lat/0206029}]; \emph{JHEP} {\bf 0406} (2004)
061 [{\tt hep-lat/0307017}].

\bibitem{KNN03}
J. Kiskis, R. Narayanan and H. Neuberger, \emph{Phys. Lett.} {\bf
B574} (2003) 65 [{\tt hep-lat/0308033}]; \\
R. Narayanan and H. Neuberger, \emph{Phys. Rev. Lett.} {\bf 91}
(2003) 081601 [{\tt hep-lat/0303023}]; \\
H. Neuberger, \emph{Phases of planar QCD on the torus}, in 
proceedings of \emph{XXIIIrd International Symposium on Lattice 
Field Theory}, PoS(LAT2005)005 [{\tt hep-lat/0509014}].

\bibitem{Teper.LAT05}
M. Teper, \emph{Large N}, in proceedings of \emph{XXIIIrd 
International Symposium on Lattice Field Theory}, PoS(LAT2005)256 
[{\tt hep-lat/0509019}].

\bibitem{nEK}
T. Eguchi and H. Kawai,
\emph{Phys. Rev. Lett.}  {\bf 48} (1982) 1063; \\
S. R. Das,
\emph{Rev. Mod. Phys.}  {\bf 59} (1987) 235.

\bibitem{AO82}
A. Gonz\'alez-Arroyo and M. Okawa,
\emph{Phys. Rev. D} {\bf 27} (1983) 2397.

\end{thebibliography}
\end{document}